\pdfoutput=1
\documentclass[aps, reprint, superscriptaddress,nofootinbib, showpacs,preprintnumbers, floatfix]{revtex4-1}
\usepackage{amsmath, latexsym, amssymb, hyperref, graphicx, color, multirow,slashed}
\usepackage[table]{xcolor}
\usepackage{tabularx}
\usepackage[T1]{fontenc}
\usepackage[capitalize]{cleveref}
\definecolor{nicered}{rgb}{.7,.1,.1}
\definecolor{nicegreen}{rgb}{.2,.7,.1}
\definecolor{lightgreen}{rgb}{.6,.9,.5}
\definecolor{darkblue}{rgb}{0,0,.5}
\definecolor{el}{rgb}{.9, .8, .7}
\definecolor{mu}{rgb}{.8, .7, .8}
\hypersetup{colorlinks, citecolor=nicegreen,linkcolor=nicered, urlcolor=darkblue}
\usepackage{caption}
\usepackage{subcaption}
\usepackage[normalem]{ulem}
\usepackage{xspace}
\usepackage{slashed}
\usepackage{slashbox}
\usepackage[bottom]{footmisc}
\usepackage[section]{placeins}
\usepackage{makecell,tabularx}
\definecolor{blue(munsell)}{rgb}{0.0, 0.5, 0.69}

\newcommand{\gev}{~\text{GeV}}
\newcommand{\tev}{~\text{TeV}}
\newcommand{\mev}{~\text{MeV}}

\newcommand{\onbb}{0\nu\beta\beta}
\newcommand{\lsim}{\buildrel < \over {_\sim}}
\newcommand{\gsim}{\buildrel > \over {_\sim}}
\def\nn{\nonumber}

\begin{document}
\raggedbottom

\preprint{ACFI-T22-03}

\title{Unravelling the left-right mixing using $0\nu \beta\beta$ decay and collider probes}
\author{Gang Li}
\email{ligang@umass.edu}
\affiliation{Amherst Center for Fundamental Interactions, Department of Physics, University of Massachusetts, Amherst, MA 01003, USA.}
\author{Michael J. Ramsey-Musolf}
\email{mjrm@sjtu.edu.cn,\, mjrm@physics.umass.edu}
\affiliation{Tsung-Dao Lee Institute and School of Physics and Astronomy, Shanghai Jiao Tong University, 800 Dongchuan Road, Shanghai, 200240 China.}
\affiliation{Key Laboratory for Particle Astrophysics and Cosmology (MOE) \& Shanghai Key Laboratory for Particle Physics and Cosmology, Shanghai Jiao Tong University, Shanghai 200240, China}
\affiliation{Amherst Center for Fundamental Interactions, Department of Physics, University of Massachusetts, Amherst, MA 01003, USA.}
\affiliation{Kellogg Radiation Laboratory, California Institute of Technology, Pasadena, CA 91125 USA.}
\author{Juan Carlos Vasquez}
\email{jvasquezcarm@umass.edu}
\affiliation{Amherst Center for Fundamental Interactions, Department of Physics, University of Massachusetts, Amherst, MA 01003, USA.}


\begin{abstract}
\noindent
In the context of the minimal left-right symmetric model, we study the interplay between current and future neutrinoless double beta ($0\nu\beta\beta$) decay experiments, long-lived particle searches at the LHC main detectors ATLAS/CMS,  and the proposed far detector MATHUSLA.  The heavy Majorana neutrino can be produced in association with an electron from the decay of $W$ boson for a non-zero left-right mixing and subsequently decays into another electron with the same charge and jets. Owing to the suppression of large right-handed charged gauge boson $W_R$ mass, the heavy neutrinos could be long-lived. 
 We show that long-lived particle (LLP) searches for heavy Majorana neutrinos in the same-sign dilepton channel at the LHC can be used to extend  $W_R$ boson mass reach relative to the reach of the Keung-Senjanovic (KS) process.
Finally, we show that sensitivities of  LLP searches at the high-luminosity LHC with main detectors ATLAS/CMS are competitive with those of future $0\nu\beta\beta$ decay searches.

\end{abstract}

\maketitle
%
%
\section{Introduction}

Neutrinos are the sole Standard Model (SM) candidates for elementary particles having a Majorana mass. The corresponding Majorana mass term in the Lagrangian, which changes lepton number by two units, is given by $\mathcal{L}_{M} \supset-y_{\nu} \overline{\ell^{C}} H^{T} H \ell / \Lambda$, where $\ell$ and $H$ are the SM left-handed doublet and Higgs doublet respectively, and $\Lambda$ is an {\it a priori} unknown mass scale.  After  electroweak symmetry breaking, the neutral component of the  Higgs doublet takes the vacuum expectation value (vev) $v/\sqrt{2}$ and gives  the Majorana mass term to neutrinos  $\mathcal{L}_{M}\rightarrow -(m_{\nu}/2)\overline{\nu^{C}}\nu$, where $m_{\nu} =y_{\nu}v^2/\Lambda$.   Choosing  $y_{\nu}\sim \mathcal{O}(1)$, the experimentally observed scale of light neutrino masses points to a high energy scale (see-saw scale) of $\Lambda\gsim 10^{14}$ GeV, which is hence not in the reach of collider experiments in the high-energy frontier and other low energy searches at the intensity frontier.

The physics associated with  Majorana neutrinos may give rise to striking experimental signatures, such as the observation of neutrinoless double beta ($\onbb$) decay~\cite{Furry:1939qr}. This process can arise solely from the presence of three light Majorana neutrinos, even if $\Lambda$, the scale of new lepton number-violating physics, is too heavy to yield other experimentally accessible signals. Indeed,  
For sufficiently large $\Lambda$, the new physics contribution to the  $\onbb$-decay amplitude is proportional to $c/\Lambda^5$, where $c$ is some Yukawa or gauge coupling. 
The light neutrino contribution is instead characterized by $G_F^2m_{\beta\beta}/p^2$, where $G_F$ is the Fermi constant, $m_{\beta\beta}$ the effective Majorana mass and $p\sim 190$ MeV. It is worth noticing that current $\onbb$-decay searches are sensitive to $m_{\beta\beta}\simeq 0.1$ eV. 
For $c\sim \mathcal{O}(1)$ and $m_{\beta\beta}\simeq 0.1$ eV, both the light neutrino and the new physics contributions are comparable if $\Lambda\sim 4$ TeV~\cite{Rodejohann:2011mu}. It thus motivates us to consider  new physics with lepton number violation (LNV) at the TeV scale.  

Ongoing   $\onbb$-decay experiments~\cite{KamLAND-Zen:2016pfg,Majorana:2017csj,CUORE:2017tlq,EXO:2017poz,GERDA:2018pmc}  place stringent constraints on the parameter space of many extensions of the SM featuring LNV at the TeV scale. Future   $\onbb$-decay experiments  with enhanced sensitivity are expected to produce new results in the near future~\cite{Albert:2017hjq,Kharusi:2018eqi,Abgrall:2017syy,Armengaud:2019loe,CUPIDInterestGroup:2019inu,Paton:2019kgy}.  In addition,   if the new physics contribution with $\Lambda\sim 1$ TeV is the dominant source of the $\onbb$-decay rate, it could render the standard baryogenesis via leptogenesis mechanism at higher scales ineffective~\cite{Frere:2008ct,Deppisch:2013jxa,Deppisch:2015yqa,Deppisch:2017ecm,Harz:2021psp} 
(due to an efficient washout of lepton number above the electroweak scale).  Thus, observation of LNV processes~\cite{Keung:1983uu} at the  Large Hadron Collider (LHC) could falsify high-scale leptogenesis models.

In Ref.~\cite{Prezeau:2003xn}, the authors proposed a  chiral perturbation theory ($\chi$PT)  framework for computing the $\onbb$-decay rate when the LNV interactions are associated with sufficiently heavy mass scales so that the new heavy states could be reliably integrated out.  For more recent developments see Refs.~\cite{Graesser:2016bpz,Cirigliano:2017djv,Cirigliano:2018yza}. For a $\chi$PT formalism applicable when the LNV mass states cannot be integrated out (with new particle masses below or at the hadronic scale), see  Ref.~\cite{Dekens:2020ttz}. From the symmetries of quark operators arising from LNV physics at the TeV scale, this framework allows a systematic classification of the corresponding effective, hadronic operators relevant to processes at the nuclear scale.  These hadronic operators can then be classified according to their chiral symmetry transformation properties and the order in which they appear in a chiral expansion.  

As an emblematic example of a well-motivated model of LNV at the TeV scale, in Ref.~\cite{Li:2020flq}, we studied the minimal left-right symmetric model (mLRSM). In general, the left- and right-handed charged gauge bosons ($W_L$ and $W_R$, respectively) may mix to form the mass eigenstates, $W_{1,2}$, with $W_1$ being the experimentally observed $W$-boson. 
 In this context, we computed in the $\chi$PT framework the leading order,  ``long-range'' contribution to the $\onbb$-decay amplitude, proportional to the quantity $\sin\xi$ (defined below) that governs $W_L$-$W_R$ mixing.
We showed that for $\sin\xi\not=0$,  the long-range contribution might dominate the $\onbb$-decay amplitude over the other contributions~\cite{Tello:2010am}.

 As emphasized in Ref.~\cite{Li:2020flq}, without the long-range contribution to the $\onbb$ decay rate,  most of the mLRSM parameter space would be inaccessible to ton-scale $\onbb$-decay experiments if cosmological data push the bound on the sum of neutrino masses to  $\sum m_{\nu}\sim 0.1$ eV. On the other hand,  even if future cosmological probes~\cite{Hazumi:2012gjy,Abazajian:2019eic,Levi:2019ggs,Scaramella:2015rra} when combined with global fits~\cite{Capozzi:2017ipn,Capozzi:2020qhw}  to neutrino oscillation data would make the light neutrino contribution to the $\onbb$-decay rate unobservable, there are still good prospects of observing a $\onbb$ decay signal in the context of the mLRSM with non-vanishing $\sin\xi$.

The experimental observation of $\onbb$ decay would not by itself give any information about the underlying new physics model.
Thus, it is essential to consider other experimental handles to distinguish between different models of LNV at the TeV scale. Previous work~\cite{Tello:2010am} considered the interplay between the $\onbb$ decay and the production of two same-sign leptons and two jets from heavy neutrinos and one on-shell right-handed $W_R$ boson (KS process~\cite{Keung:1983uu}).

In this work, we study instead the interplay between the new leading $\onbb$-decay contribution of Ref.~\cite{Li:2020flq} (shown in Fig.~\ref{fig:mLRSM_LR}) and the production of two same-sign electrons\footnote{In this work, electron denotes $e^+$ or $e^-$, thus same-sign electrons can be $e^+e^+$ or $e^-e^-$. } and jets at the LHC (shown in Fig.~\ref{fig:LHC_signal_process}). The channel at the LHC proceeds via one on-shell  $W$ boson and an on-shell heavy neutrino $N$ as intermediate states.  Both the amplitudes for the processes in  Fig.~\ref{fig:mLRSM_LR} and Fig.~\ref{fig:LHC_signal_process} are proportional to $\sin\xi$. The enhancement due to the on-shell production of the $W$ boson in comparison with $W_R$ production in the KS process compensates for the $\sin\xi$ suppression. This enhancement makes the amplitude for the process in Fig.~\ref{fig:LHC_signal_process} comparable to the amplitude for the KS process~\cite{Keung:1983uu} -- for  recent works, see Ref.~\cite{Chen:2013foz,Nemevsek:2018bbt}. 

Interestingly, the portion of the mLRSM parameter space accessible to current and future $\onbb$-decay searches has significant overlap with the parameter space leading to the production of heavy neutrinos at the LHC  with macroscopic decay length.  The region of interest corresponds to heavy neutrino masses $m_N < M_W$, where $M_W$ is the experimentally observed $W$-boson mass.  In this case, the signal would feature one prompt electron and one displaced electron with displaced jets~\cite{ATLAS:2015xit,ATLAS:2015oan}.  
 Since the heavy neutrino is produced from the on-shell $W$ boson, the electron coming from the decay of the heavy neutrino is likely to have transverse momentum falling below the lepton isolation and is thus undetected.  With these considerations in mind, we propose here two search strategies, applicable for the following situations: (1) the final state contains two same-sign electrons and at least one jet,
and hence the LNV is manifest; (2) the final state contains at least one electron and one jet in the final state. Although the latter signal does not feature LNV,  it can extend the mass reach at the LHC.  

We show that, after improvements on the current experimental analysis~\cite{CMS:2018jxx} proposed in Ref.~\cite{Cottin:2018nms}, long-lived particle (LLP)  searches at the high-luminosity LHC (HL-LHC) with an integrated luminosity of $3~\text{ab}^{-1}$ can compete with current (future) $\onbb$ decay searches in probing the parameter space of the mLRSM with an overall selection efficiency of $4.88\%$ ($30\%$).  For the maximal value of the  $\sin\xi$ and heavy neutrino masses below the electroweak scale, the new LNV search strategy enlarges the $W_R$ boson mass reach at the HL-LHC up to  $\sim  8$~TeV -- using the efficiencies reported in Ref.~\cite{Cottin:2018nms}.  This new search significantly extends current  experimental  limits set by the ATLAS Collaboration, which excludes the mass of right-handed $W$ boson below $M_{W_R} = 3.8-5$ TeV and heavy neutrino mass below  $m_N=0.1-1.8$ TeV, respectively~\cite{ATLAS:2019isd}.   We also study the displaced-vertex signal at the MATHUSLA detector~\cite{Curtin:2018mvb} and find that both charged lepton plus missing energy and $\onbb$-decay searches rule out the mLRSM model as a candidate for a LLP signal at MATHUSLA. 

Our discussion of this analysis and proposal is organized as follows. In Sec.~\ref{secII}, we review the relevant interactions in the mLRSM and the $\onbb$-decay formalism used. Later, in Sec.~\ref{secIII}, we first discuss some analytic formulas for the production and displaced decay of heavy neutrinos at the LHC main detectors ATLAS/CMS and MATHUSLA. Then, we show our proposed displaced-vertex search strategy at the LHC. In Sec.~\ref{secIV}, we present and discuss the projected sensitivity at the HL-LHC, together with the current and future $\onbb$-decay constraints. Finally, in Sec.~\ref{secV} we give our conclusions. 
%
\section{The minimal left-right symmetric model}
\label{secII}

\paragraph{The model. }The minimal left-right symmetric model~\cite{Pati:1974yy,Mohapatra:1974gc,Senjanovic:1975rk} extends the Standard Model (SM) gauge group to  $SU(3)_C\times SU(2)_L\times SU(2)_R \times  U(1)_{B-L}$, where $B$ and $L$ denote the SM abelian baryon and lepton quantum numbers. The Higgs sector is composed of two scalar triplets $\Delta_L\in (1,3,2)$, $\Delta_R\in(3,1,2)$ and one bidoublet $\Phi\in(2,2,0)$ with $(X,Y,Z)$ denoting the representations under the SU(2)$_{R,L}$ and U(1)$_{B-L}$ groups, which are given by%
 \begin{align}
\Phi =
\left( \begin{array}{ccc}
\phi_1^0 && \phi_2^+  \\
\phi_1^- && \phi_2^{0 }\\
\end{array} \right) , \quad  \Delta_{L,R}=\left( \begin{array}{ccc}
\delta^+_{L,R}/\sqrt{2} && \delta_{L,R}^{++}  \\
\delta_{L,R}^{0} &&- \delta^+_{L,R}/\sqrt{2}  \\
\end{array} \right). \nonumber \\
\end{align}
After spontaneous symmetry breaking,  the vacuum expectation value (VEV)  of the Higgs fields take the form~\cite{Mohapatra:1980yp}
 \begin{eqnarray}
\langle \Phi \rangle =
\left( \begin{array}{ccc}
v_1 && 0  \\
0 && v_2 e^{i\alpha} \\
\end{array} \right),
\end{eqnarray}
 \begin{eqnarray}
\langle \Delta_{R} \rangle=\left( \begin{array}{ccc}
0&&0 \\
v_R  &&0  \\
\end{array} \right)  , \quad  \langle \Delta_{L} \rangle=\left( \begin{array}{ccc}
0&&0 \\
v_Le^{i\theta_L} &&0  \\
\end{array} \right),
\end{eqnarray}
 where  $\alpha$ and $\theta_L$ are the spontaneous CP  phase and  $v_L \ll   v_1^2+v_2^2 \ll   v_R^2$. All the physical effects due to $\theta_L$  can be neglected, since this phase is always accompanied by the small $v_L$.
\begin{figure}
 \centering
 \captionsetup[subfigure]{justification=centering}
\begin{subfigure}[b]{0.25\textwidth}
         \centering
         \includegraphics[width=\textwidth]{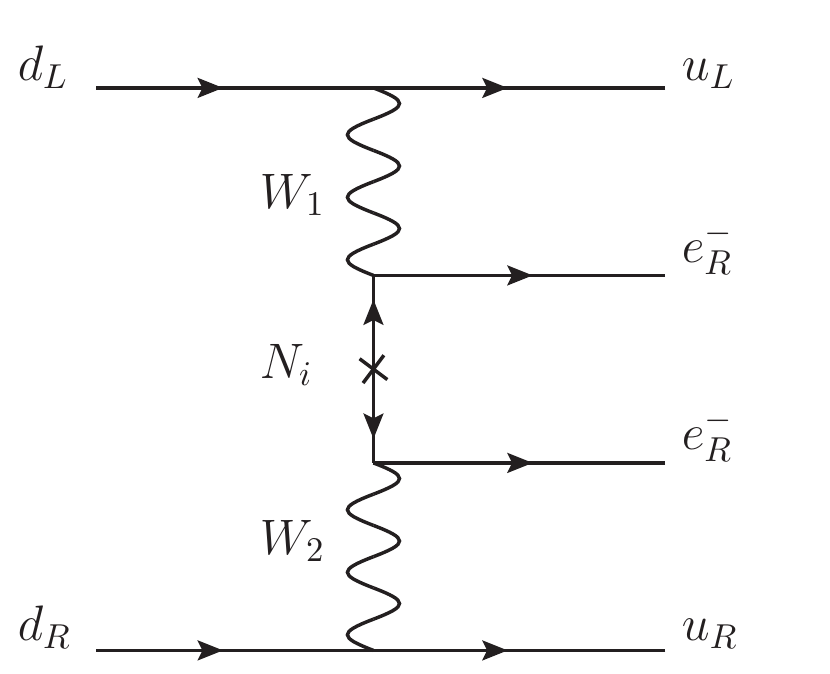}
         \caption{$\onbb$ decay}
         \label{fig:mLRSM_LR}
     \end{subfigure}
     \hfill
     \begin{subfigure}[b]{0.35\textwidth}
         \centering
         \includegraphics[width=\textwidth]{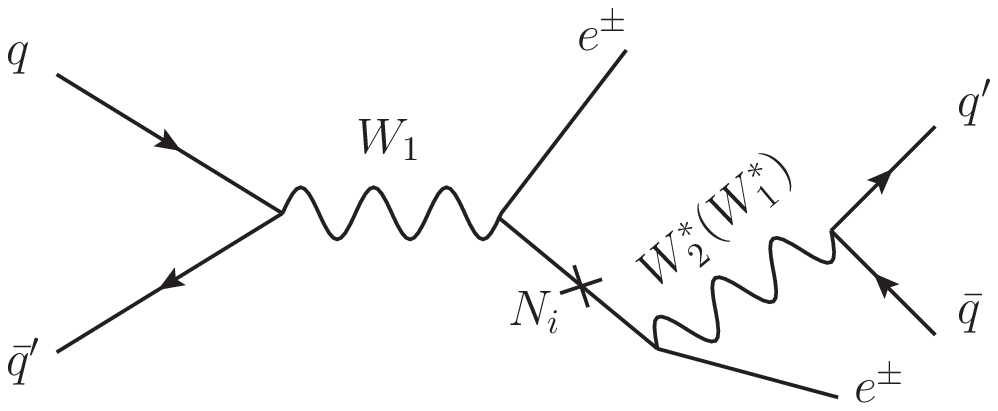}
         \caption{Signal process at the LHC}
         \label{fig:LHC_signal_process}
     \end{subfigure}
 \caption{(a): The contribution to the $\onbb$ decay arising from the left-right mixing ($\sin\xi$). Additional contributions are included in Eq.~\eqref{mNeff}. (b): The signal process at the LHC with two same-sign electrons and jets in the final state from the decay of $W_1$. Here, $N_i$ ($i=1,2,3$) denote the heavy Majorana neutrinos.}
 \label{LR_diagram}
\end{figure}
The charged current interactions are described as 

\begin{align} 
\mathcal{L}^{{\rm CC}}_{{\rm quark}} &=
\frac{g}{\sqrt{2}}\, \sum_{i,j=1}^{3} \Big[ \bar{u}_{L i} \slashed{W}_L^+ V^{\text{CKM}}_{Lij} d_{Lj}\nn\\
&\qquad +\bar{u}_{R i} \slashed{W}_R^+ V^{\text{CKM}}_{Rij} d_{R j}\Big] +  \text { h.c. }\, ,\\
\label{eq:Lag-lepton}
\mathcal{L}^{{\rm CC}}_{{\rm lepton}} &=
\frac{g}{\sqrt{2}}\, \sum_{\ell=e,\mu,\tau} \sum_{i=1}^{3} \Big[ \bar{\ell}_{L}  \slashed{W}_{L}^- V_{L\ell i} \nu_{iL} \nn\\
&\qquad + \bar{\ell}_{R}  \slashed{W}_{L}^- V_{R\ell i} N_{i}\Big] +  \text { h.c. }\, ,
\end{align}
where $g$ is the $SU(2)_{L,R}$ gauge coupling, $V_{L,R}^{\text{CKM}}$ and $V_{L,R}\equiv V_{L,R}^{\text{PMNS}}$ are the Cabibo-Kobayashi-Maskawa (CKM) and Pontecorvo-Maki-Nakagawa-Sakata (PMNS) matrices and their right-handed analogues, respectively, and ``$\text{h.c.}$'' denotes the Hermitian conjugation. As studied in Refs.~\cite{Senjanovic:2014pva,Senjanovic:2015yea}, $V_{R}^{\text{CKM}}$ is close to $V_{L}^{\text{CKM}}$ up to a small correction, which can be neglected here.

For a non-zero VEV $v_2$, the SM $W_L$ boson mixes with its heavier right-handed partner $W_R$ -- the left-right mixing. One can then  express them in terms of the light and heavy mass eigenstates, $W_1$ and $W_2$, respectively, such that
\begin{align}
    W_{L}^{+\mu} &= \cos \xi\, W_{1}^{+\mu}-\sin \xi\, \,e^{- i \alpha} \,W_{2}^{+\mu}\, ,   \\
    W_{R}^{+\mu} & = \cos \xi\, W_{2}^{+\mu}+\sin \xi\, \,e^{ i \alpha} \,W_{1}^{+\mu}\,\label{LR-mixing}.
\end{align}
The left-right mixing parameter $\xi$ is defined as $\sin\xi=\lambda\sin(2\beta)$ with the ratio of VEVs $\tan\beta=v_2/v_1$ and the ratio of masses $\lambda\simeq M_W^2/M_{W_R}^2$ where $M_W\equiv M_{W_1}$ is the mass of experimentally observed $W$-boson ($W_1$), and $M_{W_R}\simeq M_{W_2}$ is the mass parameter of $W_R$.
%
\begin{figure}
 \centering
 \includegraphics[width=0.7\columnwidth]{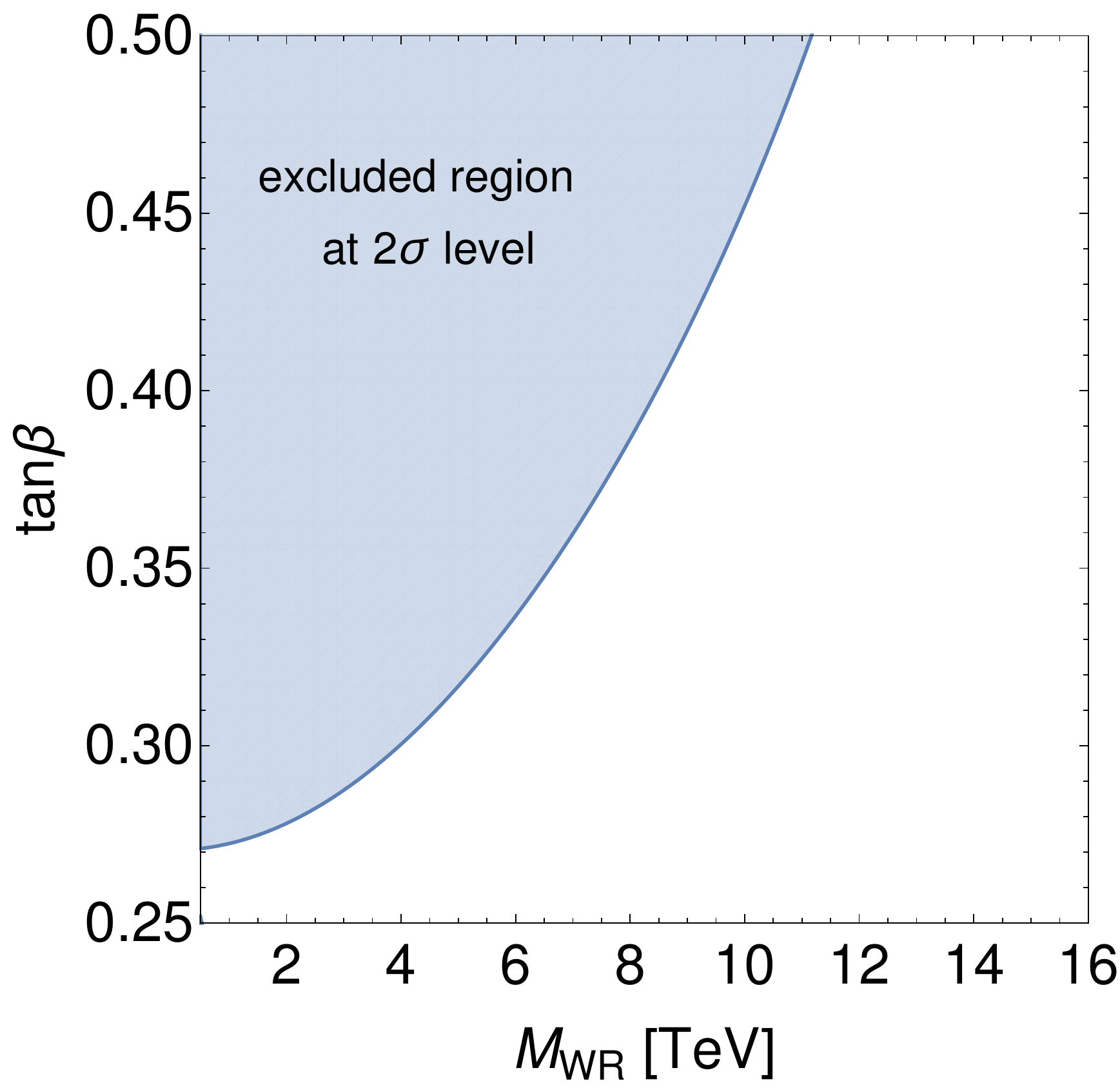} 
 \caption{  
Region (blue) excluded at $2\sigma$ level in the $M_{W_R}-\tan\beta$ plane by the $\rho$ parameter measurement $\rho = 1.00038\pm 0.00020$~\cite{Zyla:2020zbs}. 
}
 \label{MWR_tb_fig}
\end{figure}

In this work we will study signals of the left-right mixing in $\onbb$-decay experiments and at the LHC as depicted in Fig.~\ref{LR_diagram}. For the sake of illustration, we assume that the heavy Majorana neutrino $N\equiv N_1$, and $V_{Re1}=1$, and the other two ($N_2$ and $N_3$) are heavy enough and decouple.

The search for $W_R$ decaying into a high-momentum heavy neutrino and a charged lepton~\cite{ATLAS:2019isd} can exclude $W_R$ mass below 4.8~TeV in the electron channel~\footnote{It is noted that the lower limit is $M_{W_R}>3.8\tev$ if $m_N < 100\gev$ in this process, while a stronger constraint comes from the lepton plus missing energy search implying $M_{W_R} > 5\tev$ for $m_N < 40\gev$~\cite{Nemevsek:2018bbt}.}, which implies that $\lambda\lsim 2.8\times 10^{-4}$.  Note that we use the notation ``$W_R$'', which is actually ``$W_2$'', for collider searches hereafter in the usual fashion. The mass of $W_R$ and the left-right mixing parameter can also be constrained indirectly in the SM measurements. 
In this work, we revisit the constraint from the electroweak precision tests. Following Refs.~\cite{Czakon:1999ga}, the $\rho$ parameter in the mLRSM  is expressed as $\rho = 1 + [\sin^2(2\beta) +  (1-\tan^2\theta_W)^2/2]\lambda$~\cite{Czakon:1999ga}. Here, $\theta_W$ is the weak mixing angle.
The fitted value of $\rho$ parameter is $\rho = 1.00038\pm 0.00020$~\cite{Zyla:2020zbs}.  The excluded region at $2\sigma$ level in the $M_{W_R}-\tan\beta$ plane is shown in blue in Fig.~\ref{MWR_tb_fig}. As we can see, for $M_{W_R}= 5\tev$,  $\tan\beta \lesssim 0.32$ and  $\xi \lesssim 1.5\times 10^{-4}$. For a heavier $W_R$, a smaller $\xi$ while a larger $\tan\beta$ are allowed. The constraint  from the super-allowed nuclear $\beta$ decays \cite{Seng:2018yzq,Zyla:2020zbs,Cirigliano:2013xha,Gonzalez-Alonso:2018omy} is $\xi \cos\alpha \leq 1.25\times 10^{-3}$. 
Theoretically, the heavy-doublet Yukawa coupling $m_t/[v\cos (2\beta)] < \pi$ is required in order to ensure the perturbativity, so that $\tan\beta \lesssim 0.8$~\cite{Maiezza:2010ic,Dekens:2021bro}. We will assume that $\tan\beta\in [0,0.3]$ for the mass region of $M_{W_R}$ considered in this work. 

\paragraph{The effective and chiral Lagrangians.}
In  this part we present the expressions needed to evaluate the $\onbb$ decay rate within the mLRSM, which were already obtained in Ref.~\cite{Prezeau:2003xn}. The effective Lagrangian describing the $\onbb$ decay in the mLRSM below the electroweak scale is~\cite{Prezeau:2003xn,Li:2020flq}
\begin{align}
\label{Leff}
\mathcal{L}_{\text{eff}}=\dfrac{G_F^2}{\Lambda_{\beta\beta}}&\big[ C_{3R} (\mathcal{O}_{3+}^{++}-\mathcal{O}_{3-}^{++})(\bar{e}e^c -\bar{e}\gamma_5 e^c)\nonumber \\
&+ C_{3L} (\mathcal{O}_{3+}^{++}+\mathcal{O}_{3-}^{++})(\bar{e}e^c -\bar{e}\gamma_5 e^c)\nn\\
&+ C_{1} \mathcal{O}_{1+}^{++}(\bar{e}e^c -\bar{e}\gamma_5 e^c)\nn\\
&+C_{1}^\prime \mathcal{O}_{1+}^{++\prime}(\bar{e}e^c -\bar{e}\gamma_5 e^c)\big]+\text{h.c.},
\end{align}
where $G_F$ is the Fermi constant,  $1/\Lambda_{\beta\beta}=1/m_{N}$ and $m_N$ is the mass of the heavy Majorana neutrino $N$. The effective operators in Eq.~\eqref{Leff} are
\begin{align}
\mathcal{O}_{3\pm}^{++}&=(\bar{q}^\alpha_L \tau^+ \gamma^{\mu} q^\alpha_L)(\bar{q}^\beta_L \tau^+ \gamma_{\mu} q^\beta_L)~\pm \;\nn\\
&  \quad (\bar{q}^\alpha_R \tau^+ \gamma^{\mu} q^\alpha_R)(\bar{q}^\beta_R \tau^+ \gamma_{\mu} q^\beta_R),\nn\\
\mathcal{O}_{1+}^{++}&=(\bar{q}^\alpha_L \tau^+ \gamma^{\mu} q^\alpha_L)(\bar{q}^\beta_R \tau^+ \gamma_{\mu} q^\beta_R),\nn \\
\mathcal{O}_{1+}^{++^\prime}&=(\bar{q}^\alpha_L \tau^+ \gamma^{\mu} q^\beta_L)(\bar{q}^\beta_R \tau^+ \gamma_{\mu} q^\alpha_R)\;.
\end{align}
Here, $\alpha,\beta$ are the color indices, $\tau^{\pm}= (\sigma^1\pm\sigma^2)/2$ with $\sigma^1$ and $\sigma^2$ being the Pauli matrices. The Wilson coefficients $C_{3R}$, $C_{3L}$, $C_1$ are obtained by integrating out $W_1$, $W_2$ and $N_i$.  The resulting renormalization group evolution (RGE) of the Wilson coefficients  proceeds in two steps: from the scale $\mu= M_{W_2}$ to $\mu= M_{W_1}$, and from $\mu=M_{W_1}$ to $\mu=\Lambda_\mathrm{H}$ with $\Lambda_H\equiv 2\gev$. Assuming $M_{W_2}=7\tev$, we have
\begin{subequations}
\label{eq:wilson2}
\begin{align}
\begin{pmatrix}
C_1 (\Lambda_H) \\
C_1^\prime (\Lambda_H)
\end{pmatrix}
&= 
\begin{pmatrix}
0.90 & 0\\
0.48 & 2.32
\end{pmatrix}
\begin{pmatrix}
C_1(M_{W_1})\\
C_1^{\prime}(M_{W_1})
\end{pmatrix}\;,\\
C_{3L}(\Lambda_H)&= 0.81 C_{3L}(M_{W_1})\;,\\
C_{3R}(\Lambda_H)&= 0.71 C_{3R}(M_{W_2})\;,
\end{align}
\end{subequations}
where $C_1^{\prime}(M_{W_1})=0$, $C_{3L}(M_{W_1})=\xi^2$ and $C_1(M_{W_1})=-4\lambda\xi$, and $C_{3R}(M_{W_2})=-\lambda^2\left( 1+2\Lambda_{\beta\beta}^2/M_{\Delta_R}^2\right)$. The second term, which comes from the contribution of doubly charged scalar $\Delta_R^{++}$, is negligible when the left-right symmetry holds~\cite{Tello:2010am}~\footnote{When the left-right symmetry is broken, the severe constraint from flavor number violating process $\mu\to e\gamma$ gets relaxed and this term is not negligible~\cite{Dev:2018sel}.}.

From the effective Lagrangian in Eq.~\eqref{Leff}, the hadron-lepton Lagrangian valid below the chiral symmetry breaking scale is of the form~\cite{Prezeau:2003xn}
\begin{align}
\label{eq:hadron-lepton}
\mathcal{L}_{\chi\text{PT}}=&\dfrac{G_F^2F_\pi^2}{\Lambda_{\beta\beta}}\Big\{\Lambda_\chi^2 \pi^-\pi^- \bar{e}(\beta_1+\beta_2\gamma^5)e^c\nn\\
&+\partial_\mu \pi^-\partial^\mu \pi^- \bar{e}(\beta_3+\beta_4\gamma^5)e^c\nn\\
&+\Lambda_\chi/F_\pi \bar{N}i\gamma_5 \tau^+ \pi^- N  \bar{e}(\zeta_5+\zeta_6\gamma^5)e^c\nn\\
&+1/F_\pi^2 \bar{N}\tau^+N \bar{N}\tau^+N\bar{e}(\xi_1+\xi_4\gamma_5)e^c\nn\\
&+\text{h.c.}\Big\}\;,
\end{align}
with
\begin{align}
\beta_1=-\beta_2&=\ell_1^{\pi\pi} C_1 +\ell_1^{\pi\pi\prime} C_1^\prime\;,\\
\beta_3=-\beta_4&=\ell_3^{\pi\pi} (C_{3L} + C_{3R})\;,\\
\zeta_5=-\zeta_6&=\ell_3^{\pi N} (C_{3L}+C_{3R})\;,\\
\xi_{1}=-\xi_{4}&=\ell_{1}^{NN} C_1 +\ell_{1}^{NN\prime} C_1^\prime \nn\\
&+\ell_{3}^{NN} (C_{3L}+C_{3R})\;.
\end{align}
 Using the lattice calculation of $\pi^-\to \pi^+$ amplitude~\cite{Nicholson:2018mwc},  the numerical values of the low energy constants are $\ell_1^{\pi\pi}=-(0.71\pm 0.07)$, $\ell_1^{\pi\pi\prime} =-(2.98\pm 0.22)$ and $\ell_3^{\pi\pi}=0.60\pm 0.03$ in the modified minimal substraction ($\overline{\text{MS}}$) scheme at $\mu=2\gev$~\cite{Cirigliano:2018yza}. The low energy constants for the $\bar{N}N\pi \bar{e}e^c$ and the $\bar{N}N\bar{N}N \bar{e}e^c$ interactions are  estimated using naive dimensional analysis (NDA)~\cite{Manohar:1983md} with $\ell_3^{\pi N}\sim \mathcal{O}(1)$ and $\ell_{1}^{NN},\ell_{1}^{NN\prime}, \ell_{3}^{NN}\sim \mathcal{O}(1)$~\footnote{ The $\bar{N}N\bar{N}N\bar{e}e^c$ interactions are prompted as the leading-order counterterms in Ref.~\cite{Cirigliano:2018hja} with larger low energy constants being required. }.  
The chiral symmetry breaking scale $\Lambda_{\chi}=4\pi F_\pi$ and $F_\pi=91.2\mev$.

\paragraph{Neutrinoless double beta decay half-life. }The inverse half-life of $\onbb$ decay is expressed as
\begin{align}
\label{eq:half-life}
  (  T^{0\nu}_{1/2})^{-1}
 &  = \, G_{0\nu} \cdot\mathcal{M}_{\nu}^{2}\left|m_{\nu+N}^{e e}\right|^{2}
\end{align}
with $\left|m_{\nu+N}^{e e}\right|^{2}= \left|m_{\nu}^{ee}\right|^{2}+\left|m_N^{ee}\right|^2$. The effective Majorana masses of the light and heavy neutrinos are given by 
\begin{align}
m_{\nu}^{ee}   &  = \sum_{i=1}^{3} V_{Lei}^2 m_{i}(1+\ell_{\nu}^{NN} \delta^{\nu}_{NN})\;,
\end{align}
and
\begin{align}
\label{mNeff}
&|m_{N}^{ee}|^2  =  \dfrac{\Lambda_\chi^4}{72\Lambda_{\beta\beta}^2}\dfrac{\mathcal{M}_0^2}{\mathcal{M}_\nu^2}\times\bigg[(\beta_1-\zeta_5\delta_{N\pi}-\beta_3\delta_{\pi\pi}+\xi_1\delta_{NN} )^2\nn\\
&\quad +(\beta_2-\zeta_6\delta_{N\pi}-\beta_4\delta_{\pi\pi}+\xi_4\delta_{NN}  )^2\bigg]
\end{align}
with 
\begin{align}
\label{eq:delta}
\delta_{\pi\pi}&= \dfrac{2m_\pi^2}{\Lambda_\chi^2}\dfrac{\mathcal{M}_2}{\mathcal{M}_0}\;, &
\delta_{N\pi}&= \dfrac{\sqrt{2}m_\pi^2}{g_A \Lambda_\chi M}\dfrac{\mathcal{M}_1}{\mathcal{M}_0}\;, \nn\\
\delta_{NN}^{\nu}&=\dfrac{2m_\pi^2 }{g_A^2 \Lambda_\chi^2} \dfrac{\mathcal{M}_{NN}}{\mathcal{M}_\nu}\;, &
\delta_{NN}&=\dfrac{12m_\pi^2}{g_A^2 \Lambda_\chi^2}\dfrac{\mathcal{M}_{NN}}{\mathcal{M}_0}\;.
\end{align}
Here, $M$ is the nucleon mass, and $g_A=1.27$. 
The $\onbb$-decay experiments in  $^{136}$Xe are considered. The phase space factor is $G_{0\nu}^{-1}=7.11\times 10^{24}~\text{eV}^2\cdot\text{yr}$~\cite{Kotila:2012zza,Stoica:2013lka}.
We use the nuclear matrix elements calculated in quasiparticle random phase approximation (QRPA)~\cite{Hyvarinen:2015bda} and shell model~\cite{Horoi:2017gmj,Menendez:2017fdf}  methods, which are tabulated Tab.~\ref{tab:NMEs}.

\begin{table}[]
    \centering
\begin{tabular}{c|cccccc}
   & $\mathcal{M}_\nu$ & $\mathcal{M}_0$ & $\mathcal{M}_1$ & $\mathcal{M}_{2}$ & $ \mathcal{M}_{NN}$ \\ 
\hline
\hline
QRPA~\cite{Hyvarinen:2015bda}          & $-2.85$   & $-2.64$  & $-5.58$    & $-4.26$ & $-1.53$  
\\
shell~\cite{Horoi:2017gmj}          & $-1.99$   &    $-1.11$ & $-2.06$ & $-1.50$ & $-0.92$ 
\\
shell~\cite{Menendez:2017fdf}         & $-2.31$  & $-1.50$   & $-2.91$    & $-2.16$ & $-1.28$ 
\end{tabular}
    \caption{Nuclear matrix elements in QRPA and shell model methods.}
    \label{tab:NMEs}
\end{table}

Notice the chiral suppression $\sim m_{\pi}^2/\Lambda_{\chi}^2, m_{\pi}^2/(\Lambda_{\chi} M)$ in Eq.~\eqref{eq:delta}. 
The contributions proportional to $\beta_1$ and $\beta_2$ in Eq.~\eqref{mNeff} give the leading contribution to the $\onbb$-decay rate for $\tan\beta\gtrsim 0.1$~\cite{Li:2021fvw}. To make the interplay of $\onbb$-decay and collider searches as transparent as possible, we have assumed that only one heavy Majorana neutrino $N$ contributes. At the same time, the other two are heavy enough and decouple. The general case with three heavy neutrinos is a straightforward generalization of the aforementioned simple case.  
If one (or more) of the other heavy neutrinos, say $N_2$, also gives a non-negligible contribution to $\onbb$ decay, the heavy neutrino effective Majorana mass 
$m_N^{ee}\propto V_{Re1}^2 1/m_N + V_{Re2}^2 1/m_{N_2}$.  These two terms might interfere destructively for specific $V_R$.
The cancellation between contributions from the exchange of different heavy neutrinos
would possibly lead to a largely suppressed $\onbb$-decay rate, whose implication we will return to in Sec.~\ref{secV}.

\section{Collider probes of the left-right mixing}
\label{secIII}
This section discusses the experimental setup and explains our strategies for the LLP searches at the LHC main detectors ATLAS/CMS and the far detector MATHUSLA.  As we shall see,  depending on whether the signal shown in Fig.~\ref{fig:LHC_signal_process} exhibits explicit LNV in the form of two same-sign electrons or not,  we propose two search strategies at the LHC. One search strategy corresponds to the   LNV signal region, requiring two same-sign electrons and at least one jet in the final state. The other search strategy corresponds to the lepton-number-conserving (LNC) signal region, requiring at least one electron and one jet in the final state. We study and estimate the expected mass reach for both signal regions at the HL-LHC.

\begin{figure}
 \centering
 \includegraphics[width=0.9\columnwidth]{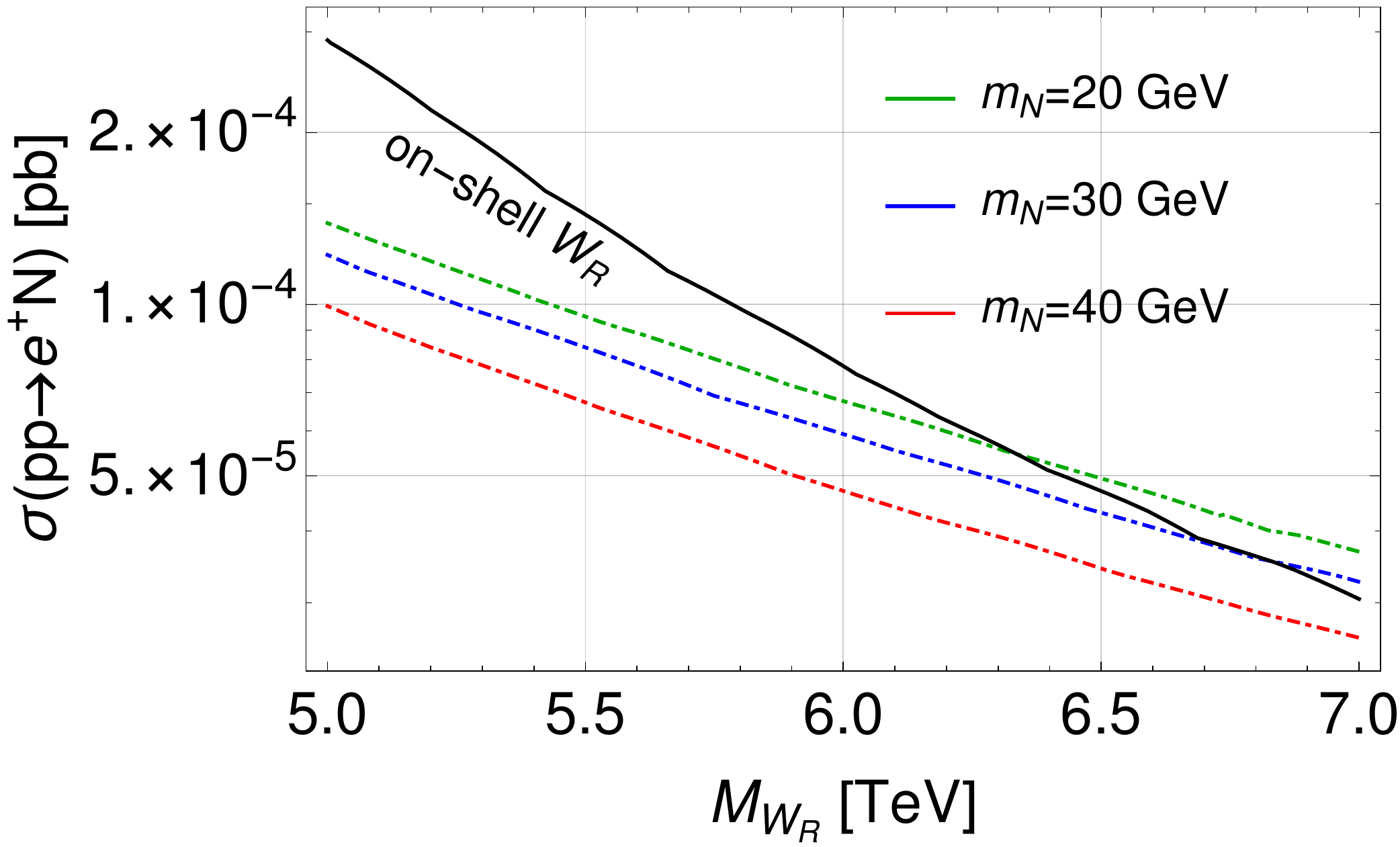} 
 \caption{  
Cross sections  for the processes  $pp\rightarrow W^+_R\rightarrow e^+ N$ and $pp\rightarrow W^+\rightarrow e^+ N$ at the 13~TeV LHC a function of $M_{W_R}$. For the former process (black curve), the heavy neutrino mass $m_N=20\gev$ is assumed. For the latter process as shown in Fig.~\ref{fig:mLRSM_LR} with $\tan\beta = 0.25$, the heavy neutrino mass is assumed to be $m_N= 20,30,40$ GeV for green, blue and red curves, respectively. 
 }
 \label{cross_sections}
\end{figure}

In Fig.~\ref{cross_sections}, we compare the cross sections of the processes  $pp\rightarrow W^+_R\rightarrow e^+ N$ and $pp\rightarrow W^+\rightarrow e^+ N$ at the LHC with the center-of-mass energy $\sqrt{s}=13\tev$, the latter of which is shown in Fig.~\ref{fig:LHC_signal_process}.
The black curve corresponds to the cross section $\sigma(pp\rightarrow W^+_R\rightarrow e^+N)$ for  $m_N=20$~GeV. As $m_N\ll M_{W_R}$,  $\sigma(pp\rightarrow W^+_R\rightarrow e^+N)$ is independent of $m_N$ to a very good approximation.  The green, blue and red curves correspond to the cross sections $\sigma(pp\rightarrow W^+\rightarrow\ell^+N)$ for $m_N= 20,30,40$~GeV from upper to lower, respectively, with the ratio of VEVs $\tan\beta=0.25$.
From Fig.~\ref{cross_sections},  we see that $\sigma(pp\rightarrow W^+\rightarrow e^+N)$ is possibly larger than  $\sigma(pp\rightarrow W^+_R\rightarrow e^+N)$ for $M_{W_R}\gtrsim 6.3 \tev$. Hence, for a non-zero $W_L-W_R$ mixing and sufficiently large $M_{W_R}$,  the process $pp\rightarrow W^+\rightarrow e^+N$  becomes the dominant production channel for the production of heavy neutrino $N$ at the LHC. The main advantage of this channel is that the $W_R$ is never produced as an on-shell particle, thus there is no phase-space suppression when $W_R$ is heavy. The reduction of $\sigma(pp\rightarrow W^+\rightarrow e^+N)$ as  $M_{W_R}$ increases comes from the dependence of the cross-section on the $W_L-W_R$ mixing $\xi$ shown in Eq.~\eqref{LR-mixing}.    

%

%
\subsection{Long-lived heavy neutrinos at the LHC: analytical formulae}

Here, we will present the collider study of the process shown in Fig.~\ref{LR_diagram}b, namely the on-shell production of experimentally observed $W$-boson and, via the left-right mixing, its decay into an electron and a heavy neutrino $N$ in the process $pp\rightarrow W^{\pm}\rightarrow e^{\pm}N$. The heavy neutrino $N$ subsequently decays into one electron and two jets. 

As with the $\onbb$-decay study, in this work, we assume one heavy neutrino $N_1=N$ (the other two are considered heavy, so they decouple) and the right-handed mixing $V_{R e1} = 1$.  Then the heavy neutrino $N$ mainly decays into an additional electron and two jets. The final state can therefore include two same-sign electrons and jets. For heavy neutrino mass below the electroweak scale, its decay products can have a macroscopic decay length, such that the final signal features one prompt electron and one displaced electron in association with displaced jets.  Below we will discuss the relevant portion of the parameter space leading to displaced-vertex signals at the LHC main detectors ATLAS/CMS and the  MATHUSLA~\cite{Curtin:2018mvb} detector.

The cross section of the process $pp\to W^\pm \to e^\pm N$ is  expressed as
\begin{align}
\sigma_{eN}= \sigma(pp\rightarrow W^{\pm})\text{Br} (W^{\pm}\rightarrow e^{\pm}N)\;,
\end{align}
where  $ \sigma(pp\rightarrow W^{\pm})$ is the $W$ boson production cross section at the LHC, and ``Br'' denotes the branching ratio. The partial decay width of $W^{\pm}\rightarrow e^{\pm}N$ is given by
\begin{align}
\Gamma (W^{\pm}\rightarrow e^{\pm}N) =& \xi^2 g^2 
\frac{M_W}{48\pi} \left( 1-\frac{m_N^2}{M_W^2}\right)^2 \nn\\
&\times \left(1+\frac{1}{2}\frac{m_N^2}{M_W^2}	\right).
\end{align}

In the mLRSM, the total width of the heavy neutrino $N$ is dominated by the decay into two electrons and two jets through an off-shell $W_R$ boson or an off-shell $W$ boson if the left-right mixing is allowed.\footnote{We compute the decay width and agree with the total width reported in Ref.~\cite{Helo:2013esa}.}
The decay branching ratio of $N\to e^+ jj$ or $N\to e^- jj$ is expressed as
\begin{align}
\text{Br}_{ejj} = \dfrac{1}{2}\ \dfrac{0.7 \left[1+\sin^2 (2\beta)\right]}{0.7 +  \sin^2 (2\beta)}\;,
\end{align}
which equal to 0.45 for $\tan\beta=0.3$.

The probability of $N$ decaying inside a detector is expressed as
\begin{align}
\label{eq:prob}
P_{\text{decay}}(d; L_1,L_2) 
=  e^{-L_1/d}-e^{-L_2/d}\;,
\end{align}
where $d=bc\tau $ is the  decay length of $N$ in the lab frame with the boost factor $b=M_W/(2m_N)$, and $L_1$ and $L_2$ $(L_1<L_2)$ are the distances from the  interaction point where the LLP enters and exits the decay volume inside the detector. 

Following Ref.~\cite{Li:2021fvw}, we use the following analytical formula to approximate  the observed number of events from a long-lived $N$ at the LHC main detectors ATLAS/CMS (marked with the superscript ``LHC''):
\begin{align}
\label{eq:obs_LHC}
N_{\text{obs}}^{\text{LHC}} = \sigma_{eN}\  \text{Br}_{ejj} \  \mathcal{L}\ \epsilon_{\text{LLP}}^{\text{LHC}}\ \epsilon_{\text{prompt}}^{\text{LHC}}\ P_{\text{decay}}\;,
\end{align}
where  $L_1=5~\text{mm}$ and $L_2=300~\text{mm}$~\cite{ATLAS:2015oan} for the inner detector are considered. $\mathcal{L}$ is the LHC integrated luminosity, and $\epsilon_{\text{prompt}}^{\text{LHC}}$ and $\epsilon_{\text{LLP}}^{\text{LHC}}$ denote the efficiencies of the trigger/identification of the prompt electron and selecting the long-lived $N$ at the LHC main detectors ATLAS/CMS, respectively.

At the far detector MATHUSLA (marked with the superscript ``MATH''), the following formula is proposed in Ref.~\cite{Curtin:2018mvb},
\begin{align}
\label{NMATHUSLA}
N_{\text{obs}}^{\text{MATH}} &= \sigma_{eN}\  \text{Br}_{ejj}  \ \mathcal{L} \ \epsilon_{\text{LLP}}^{\text{MATH}}\ \epsilon_{\text{geometric}}\ P_{\text{decay}}\;.
\end{align}
where the geometric acceptance $\epsilon_{\text{geometric}}=0.05$ describing the fraction of long-lived $N$ traversing the MATHUSLA detector, which has the size of $L_1=200$~m and $L_2=230$~m~\cite{Curtin:2018mvb}\footnote{Recently, there is upgrade for MATHUSLA~\cite{MATHUSLA:2020uve}, which claims to give very similar LLP sensitivities. }, and $\epsilon_{\text{LLP}}^{\text{MATH}}$ is the efficiency of detecting the long-lived $N$. 

As we will show in Sec.~\ref{secV}, the analytical formulae can describe the sensitivities to long-lived $N$ at the LHC well compared to the detailed collider simulation in the following subsection.

\subsection{Detailed collider simulation}
%
\begin{table*}
\centering
\begin{tabular}{ r | c  }
\hline
\hline
 \multicolumn{2}{c}{  \cellcolor{mu} Cut flow in the LNV signal region}  \\
 \hline 
          \centering Selection Cut   &   Description\\
\hline
\hline
    $e^{\pm} e^{\pm} j$, no $b$ jets  & Signal selection. Reduces $W3j$, $Z3j$ backgrounds  \\
    $\slashed{E}_T\equiv |\vec{\slashed{p}}_T| <30$ GeV & Reduces $W3j$, $Wj$ backgrounds  \\
    $p_T(e_1) < 55 \text{ GeV}$ & Reduces mostly $t \overline t (j)$ background   \\
    $M_T(e_i, \slashed{E}_T) < 30\gev$ & Reduces mostly $ZZ(j)$ background   \\
    $M(e_1, e_2) < 80~ \text{GeV} $, $M(e_{i}, {\rm \slashed{E}_T}) < 60 \text{ GeV}$ & Reduces $WZj$, $ZZj$, $Z3j$ backgrounds  \\
    $l_T(e_2) > 0.5 \text{ mm}$, $ d_{xy}(e_1) > 0.02 \text{ mm} $ &   Reduces all backgrounds     \\
     \hline
     \hline
\multicolumn{2}{c}{  \cellcolor{mu} Cut flow in the LNC signal region}  \\
\hline 
\centering Selection Cut   &   Description\\
\hline
\hline
   $e^{\pm} j$, no $b$ jets  & Signal selection. Reduces  backgrounds but mostly $4j$  \\
    $\slashed{E}_T\equiv |\vec{\slashed{p}}_T| <30$ GeV & Reduces $t \overline t (j)$ background  \\
    $p_T(e_1) < 55 \text{ GeV}$ & Reduces mostly $W3j$ background   \\
    $M_T(e, \slashed{E}_T) < 30\gev$ & Reduces mostly $Z3j$, $W3j$, 4j backgrounds   \\
   $M(e, {\rm \slashed{E}_T}) < 60 \text{ GeV}$ & Reduces $t \overline t (j)$, $Z3j$, $W3j$ backgrounds  \\
  $ d_{xy}(e) > 0.1 \text{ mm} $ &   Reduces all backgrounds     \\
     \hline
\end{tabular}
\caption{Selection criteria used to reduce the SM backgrounds at the LHC for both LNV and LNC signal regions.  }
\label{tabBckgq_description}
\end{table*}

Since the heavy neutrino $N$ is produced from the decay of the on-shell $W$ boson as depicted in Fig.~\ref{LR_diagram}, the transverse momentum distribution of $N$ has the peak around $\sim (1-m_N^2/M_W^2) M_W/2$. Thus the electron from the decay of $N$ is likely to have transverse momentum falling below the lepton isolation requirement in standard searches by the ATLAS and CMS Collaborations. Therefore, the electron coming from $N$ can go undetected.
With this in mind, in this subsection, we discuss two signal regions. The first one requires two same-sign electrons and at least one jet, which we call the LNV signal region. There is another region featuring one prompt electron and a jet containing the products of $N$ decay, including the secondary electron. In this region, the lepton number may or may not be conserved. However, this region could be used with the first one to extend the exclusion and mass reach, as we shall discuss in detail in this subsection. 

We use \texttt{MadGraph5\_aMC@NLO}~\cite{Alwall:2011uj} to simulate signal events, which are passed to \texttt{Pythia8}~\cite{Sjostrand:2014zea} for hadronization and \texttt{Delphes3}~\cite{deFavereau:2013fsa} for fast detector simulation. We use the modified \texttt{Delphes} module of Ref.~\cite{Nemevsek:2018bbt} to parameterize the displacement and smearing of transverse impact parameter relative
to the primary vertex.  The \texttt{MadAnalysis5} package~\cite{Conte:2012fm} is tuned to impose cuts on the transverse distance and transverse impact parameter of final state particles. 
 The   two signal regions we consider at the LHC main detectors ATLAS/CMS consist of: 

\paragraph{Lepton-number-violating signal region.} In this signal region, we consider two same-sign electrons, which are labeled as $e_1$ and $e_2$ sorted by transverse momentum,  and at least one jet (we veto the b-jets) together with the following cuts. 
      \begin{enumerate}
    \item Missing energy $\slashed{E}_T\equiv |\vec{\slashed{p}}_T| <30$ GeV.
    \item Transverse momentum of the leading electron $p_T(e_1) < 55 \text{ GeV}$. 
    \item Transverse mass $M_T(e_i, \slashed{E}_T) < 30\gev$ for $i=1,2$, where  $M_T(e, \slashed{E}_T) \equiv \sqrt{2p_T(e)\slashed{E}_T (1-\cos\Delta\phi)}$ with  $\Delta\phi$ the azimuthal angle difference between the electron transverse momentum and missing transverse momentum. 
    \item Invariant mass $M(e_1, e_2) < 80~ \text{GeV} $ and $M(e_i, {\rm \slashed{E}_T}) < 60 \text{ GeV}$ for $i=1,2$. Here, $M(e_i, {\rm \slashed{E}_T}) = 2\big[p_T(e_i) \slashed{E}_T \cosh \Delta y - \vec{p}_T(e_i)\cdot \vec{\slashed{p}}_T \big]$ with $\Delta y$ the rapidity difference~\cite{Zyla:2020zbs}.
 
    \item Transverse distance relative to the primary vertex~\cite{ATLAS:2015oan} of the sub-leading electron  $l_T(e_2) > 0.5 \text{ mm}$ and transverse impact parameter~\cite{CMS:2018jxx} of the leading electron $ d_{xy}(e_1) > 0.02 \text{ mm} $. 
   \end{enumerate}

   %

     \paragraph{Lepton-number-conserving signal  region.} In this signal region we require at least one electron and at least one jet (we veto the b-jets), in addition with the following cuts. 
     \begin{enumerate}
    \item Missing energy $\slashed{E}_T<30$ GeV.
    \item Transverse momentum of the  electron $p_T(e) < 55 \text{ GeV}$. 
    \item Transverse mass $M_T(e, \slashed{E}_T) < 30 \text{ GeV}$. 
    \item Invariant mass $ M(e, \slashed{E}_T) < 60 \text{ GeV}$.
    \item We require for the transverse impact parameter of the electron $ d_{xy}(e) > 0.1\text{ mm}$. 
   \end{enumerate}
A similar analysis was performed by the CMS collaboration in Ref.~\cite{CMS:2018jxx} but without the displaced-vertex cut on the transverse distance. The main backgrounds for the signal regions under consideration are: $WZj$, $ZZj$, $4j$,  $t\overline{t}j$,   $W3j$, $Wj$, $Z3j$ and $tj$.    We validate our background samples against the results reported in Ref.~\cite{CMS:2018jxx} and find good agreements with the invariant mass and transverse momentum distributions. In Tab.~\ref{tabBckgq_description}, we show the cuts we used for both the LNV and the LNC signal regions. We also give a brief description for each cut highlighting their main features.

In Tab.~\ref{tabsignaleff}, we show for several benchmark values of $M_{W_R}$ and $m_N$  the signal efficiencies in both of the LNV and the LNC signal regions after the set of cuts are applied. Roughly, in the LNV signal region, we find that the signal selection efficiencies are about $(0.1-1)\%$, which provide an improvement with respect to those reported in the CMS analysis of Ref.~\cite{CMS:2018jxx}  due to the displaced-vertex cut.  In the LNC signal region,  we obtain signal selection efficiencies between $(0.7-2)\%$. 
In Appendix~\ref{appendix}, we provide selection efficiencies for the SM backgrounds, which were entirely rejected after imposing all of the cuts.

\begin{table}
\centering
\tabcolsep=0.05cm
\begin{tabular}{ r | c | c | c}
\hline
\hline
 \multicolumn{4}{c}{  \cellcolor{mu} Signal efficiencies in $\%$}  \\
 \hline 
         \backslashbox{LNV}{LNC} 
    &   $m_N=20$ GeV &  $m_N=30$ GeV &  $m_N=40$ GeV  \\
\hline
\hline
$M_{W_R}=7$ TeV  &  \backslashbox{0.77}{1.9} &  \backslashbox{0.83}{1.8} & \backslashbox{0.14}{1.8} \\ \hline
$M_{W_R}=9$ TeV  & \backslashbox{1.0}{2.1} & \backslashbox{1.0}{1.9} & \backslashbox{0.66}{2.1} \\ \hline
$M_{W_R}=10$ TeV & \backslashbox{0.72}{1.9} & \backslashbox{1.1}{1.8} & \backslashbox{0.83}{2.2}
   \\ \hline
\end{tabular}
\caption{Signal efficiencies (in unit of \%) in the LNV and  LNC signal regions after all of the selection cuts are applied. }
\label{tabsignaleff}
\end{table}

\section{Results and discussion}
\label{secIV}
\begin{figure}[!]
 \centering
 \captionsetup[subfigure]{justification=centering}
\begin{subfigure}[b]{0.3\textwidth}
         \centering
         \includegraphics[width=\textwidth]{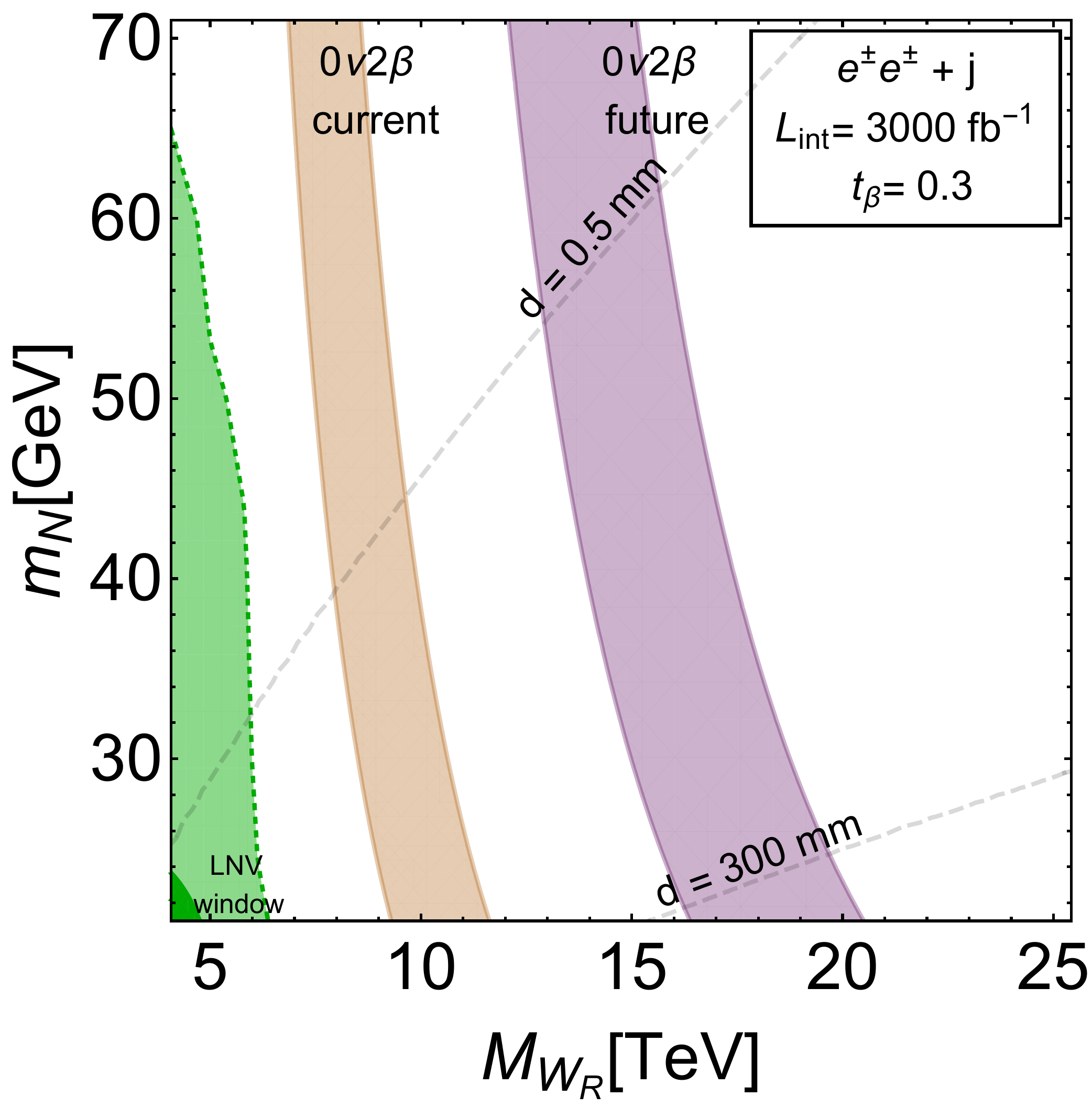}
         \caption{}
         \label{LLP_LHC_a}
     \end{subfigure}
     \hfill
     \begin{subfigure}[b]{0.32\textwidth}
         \centering
         \includegraphics[width=\textwidth]{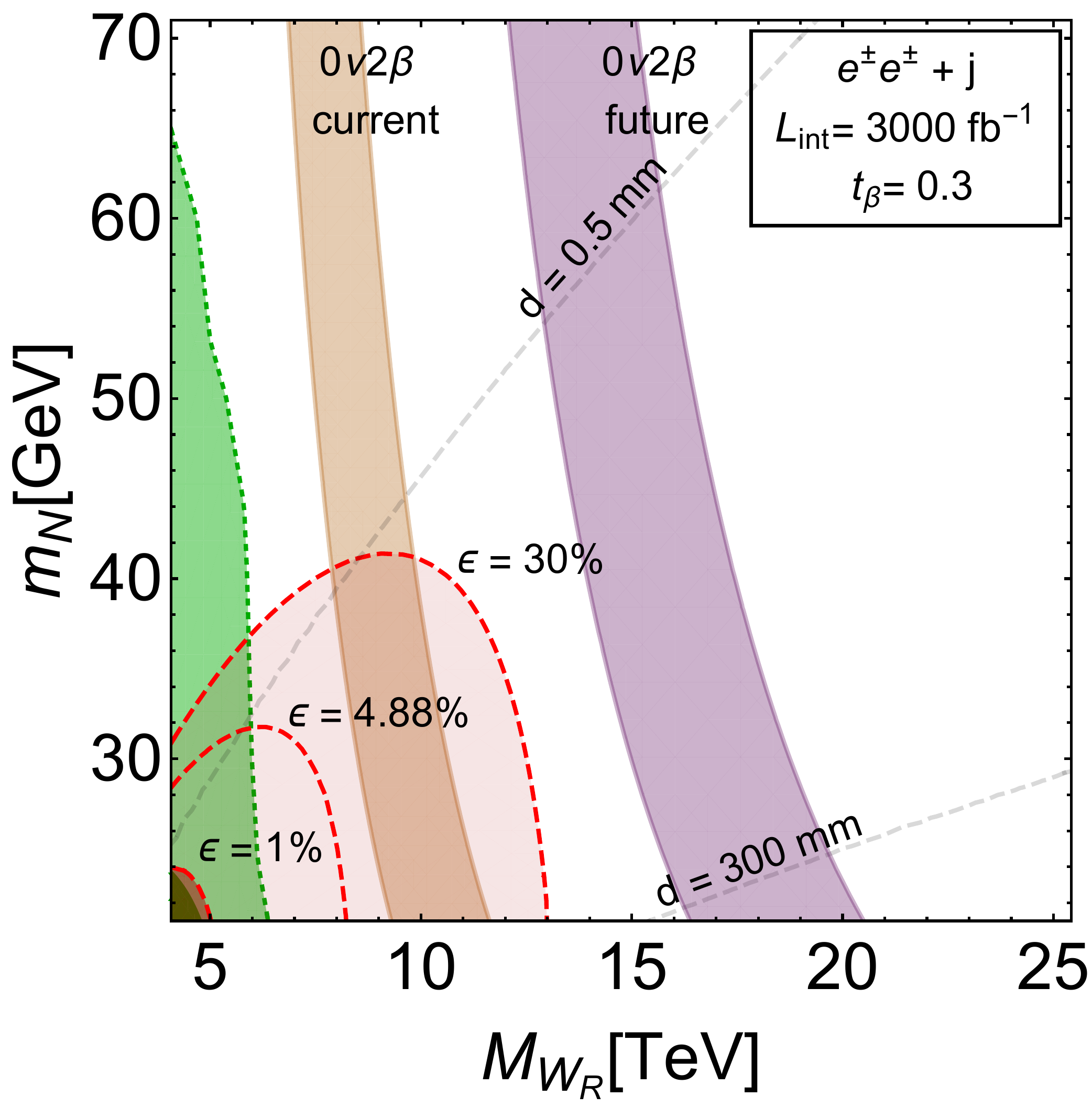}
         \caption{}
         \label{LLP_LHC_b}
     \end{subfigure}
 \caption{ 
 Panel (a) gives the mass reach of LLP searches at the HL-LHC main detectors ATLAS/CMS and $\onbb$ decay in the $M_{W_R}$-$m_N$ plane. Dark (light) green region represents the $2\sigma$ exclusion at the HL-LHC in the LNV (LNC) signal region. The area on the left of the orange (purple) band, which indicates uncertainties in NMEs, is excluded by current (future) $\onbb$-decay experiments at 90\% C.L.. Panel (b) includes the HL-LHC reach (red shaded regions) estimated using the analytical formula in Eq.~\eqref{eq:obs_LHC} with $\epsilon \equiv \epsilon_{\text{LLP}}^{\text{LHC}}\ \epsilon_{\text{prompt}}^{\text{LHC}} = (1,4.88,30)\%$. The dashed gray lines correspond to the boosted decay length of heavy neutrino.  The ``LNV window'' denotes the region in the MWR-MN plane for which two same-sign leptons plus jets signal would be observable at the LHC
 }
 \label{LLP_LHC}
\end{figure}
This section is focused on the interplay of $\onbb$ decay and LLP searches in both LNV and LNC signal regions at the HL-LHC. 
In Fig.~\ref{LLP_LHC_a}, we show the sensitivities to right-handed gauge boson mass $M_{W_R}$ and heavy neutrino mass $m_N$ for the maximal value $\tan\beta=0.3$.
The dark green area represents the reach -- $2\sigma$ exclusion -- at the HL-LHC  main detectors ATLAS/CMS in the LNV signal region with $m_N\sim (10-20)$ GeV and  $M_{W_R} \lesssim  5$ TeV. The light green area represents the reach in the LNC signal region.
In this case, the relevant heavy neutrino mass range extends over $(20-65)$~GeV, and the $W_R$ mass reach is  $\sim 6.5$~TeV.

The area on the left of the orange (purple) shaded band in Fig~\ref{LLP_LHC_a} represents the excluded parameter space from current (future) $\onbb$-decay experiments at 90\% confidence level (C.L.), respectively.   Current $\onbb$-decay searches already exclude a positive LNV signal in LLP searches at the HL-LHC with $\sim 1\%$ efficiencies assuming no cancellation between different heavy neutrino contributions to $\onbb$-decay rate in the mLRSM.  We estimate the uncertainties in  NMEs by evaluating the $\onbb$-decay rate using the  QRPA~\cite{Hyvarinen:2015bda} and shell model~\cite{Horoi:2017gmj,Menendez:2017fdf} methods, and find that the variations on the exclusion curves are within 30$\%$ in the mass range of $M_{W_R}$ and $m_N$ under consideration.  The aforementioned uncertainties are then shown in the exclusion orange and purple bands of Fig~\ref{LLP_LHC_a} for the current and future ton-scale $\onbb$-decay experiments, respectively.  
\begin{figure}
 \centering
 \includegraphics[width=0.8\columnwidth]{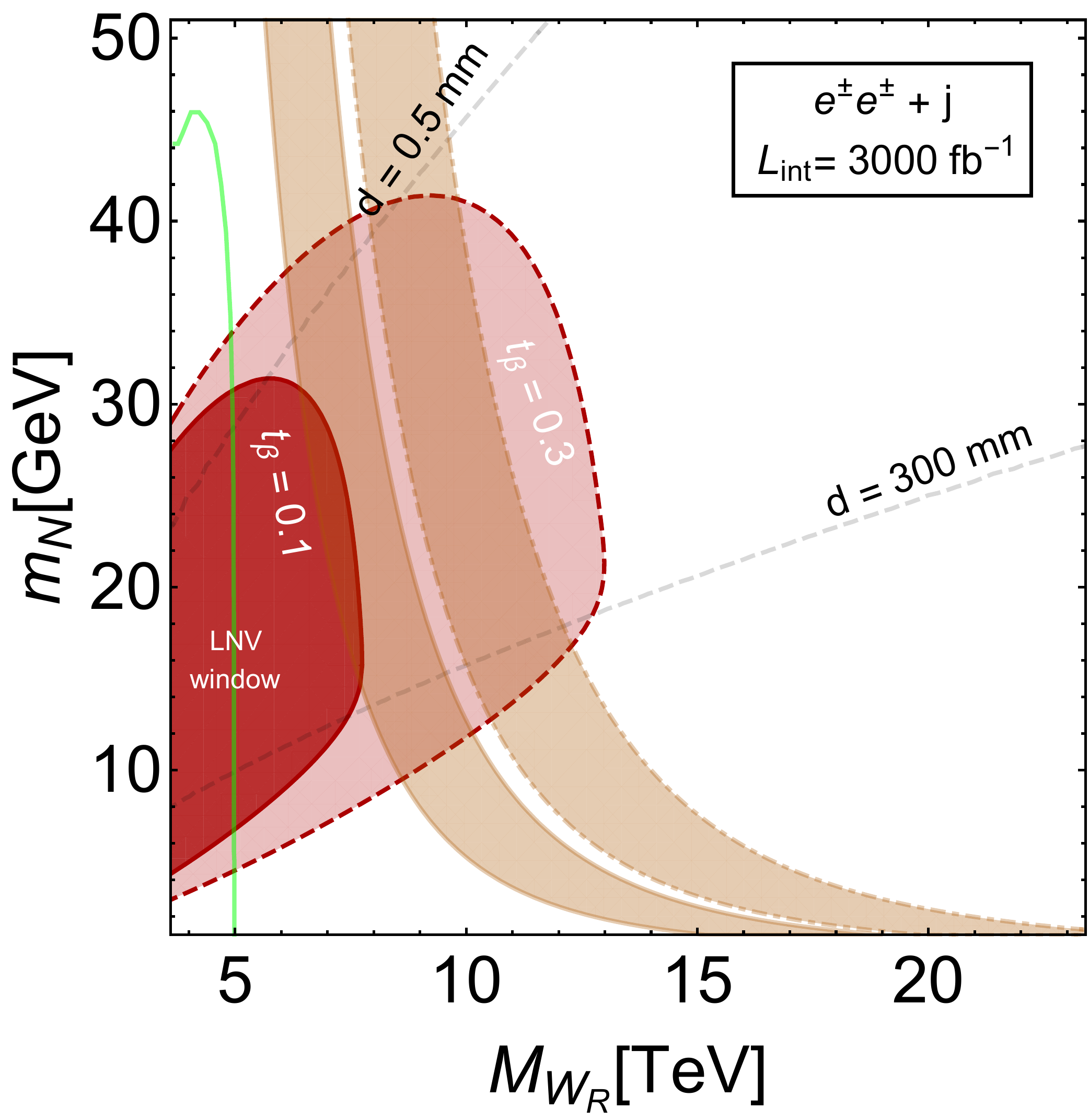} 
 \caption{Dependence on $\tan\beta$ of the sensitivities of the LLP search at the HL-LHC  main detectors ATLAS/CMS and current $\onbb$-decay searches. 
 Red shaded regions correspond to the HL-LHC reach at main detectors for $t_\beta\equiv \tan\beta =0.1, 0.3$  and the efficiency  $\epsilon \equiv \epsilon_{\text{LLP}}^{\text{LHC}}\ \epsilon_{\text{prompt}}^{\text{LHC}} =30\%$. The area on the left of the orange  band with solid (dashed) boundary is excluded by current $\onbb$-decay experiments for $\tan\beta=0.1$~(0.3). The dashed gray lines correspond to the boosted decay length of heavy neutrino. The green curve denotes  the projected exclusion limit at the LHC with 36.1 $\text{fb}^{-1}$ from the $e^{\pm}+ \slashed{E}_T$ of $W'$ search by the ATLAS Collaboration found in Ref.~\cite{Nemevsek:2018bbt}
 }
 \label{LHC_several}
\end{figure}
%

As discussed at the end of Sec.~\ref{secII}, there could be a large cancellation if two or more heavy neutrinos contribute to the $\onbb$-decay rate. Thus, even though in Fig.~\ref{LLP_LHC_a}, most of the parameter space 
accessible through an
LLP search at the HL-LHC main detectors ATLAS/CMS is excluded by current $\onbb$-decay experiments in the case of one heavy neutrino, the LLP search that we propose provides a complementary test of the same parameter space.

We also compare the HL-LHC reach  shown in Fig.~\ref{LLP_LHC_a} with the one obtained using the analytic formula shown in Eq.~\eqref{eq:obs_LHC}  by considering three representative choices of the overall efficiency  $\epsilon_{\text{LLP}}^{\text{LHC}}\ \epsilon_{\text{prompt}}^{\text{LHC}} = (1,4.88,30)\%$ with the $2\sigma$ exclusion limit given by $N_{\text{obs}}^{\text{LHC}}=3$ assuming zero background.  
We can roughly reproduce the  reach from detailed simulation (LNV signal region) with the one obtained using the analytic formula for $\epsilon_{\text{LLP}}^{\text{LHC}}\ \epsilon_{\text{prompt}}^{\text{LHC}} = 1\%$.
 In Ref.~\cite{Cottin:2018nms}, a proposal based on displaced vertices with  associated charged tracks claims the efficiency could be improved to be $\epsilon_{\text{LLP}}^{\text{LHC}}\ \epsilon_{\text{prompt}}^{\text{LHC}} = 4.88\%$. 
Armed with the analytical formula in Eq.~\eqref{eq:obs_LHC}, we obtain the sensitivity of the LLP search at the HL-LHC main detectors ATLAS/CMS  with the realistic improvement of efficiency proposed in Ref.~\cite{Cottin:2018nms}, which is illustrated in Fig.~\ref{LLP_LHC_b}. We find that the LLP search and current $\onbb$-decay searches are complementary given the uncertainties in available NMEs for $^{136}$Xe evaluated using QRPA and shell model methods and can extend the $W_R$ mass reach $(\sim 8\tev)$ with respect to the standard KS searches -- e.g., Ref.~\cite{Nemevsek:2018bbt}.

Moreover, there could be 
an improvement of the overall efficiency in the mLRSM compared to  $\epsilon_{\text{LLP}}^{\text{LHC}}\ \epsilon_{\text{prompt}}^{\text{LHC}} = 4.88\%$, which was obtained from the analysis~\cite{Cottin:2018nms} in the context of  SM with an additional heavy Majorana neutrinos, due to a larger decay length of the heavy neutrino. To understand it, we compare the decay lengths $d_{N,LR}$ of the heavy neutrinos in both scenarios~\cite{Helo:2013esa} 
\begin{align}
d_{N}&\simeq 0.37\, b\left(\frac{10~\mathrm{GeV}}{m_{N}}\right)^{5}\left(\frac{10^{-8}}{\left|V_{l 4}\right|^{2}}\right)[m]\,, \\
d_{LR}&\simeq \frac{1.2\, b}{1+ \sin^2 (2\beta)} \left(\frac{10~ \mathrm{GeV}}{m_{N}}\right)^{5}\left(\frac{M_{W_{R}}}{10 \,\mathrm{TeV}}\right)^{4}[m] \,,
\end{align}
where  $b$ is the boost factor defined in Sec.~\ref{secIV}. 
It is shown in Ref.~\cite{Cottin:2018nms} that $V_{l4}\sim 10^{-4}-10^{-5}$ in the heavy neutrino mass range  $m_N=(15-25)\gev$ can be probed at the HL-LHC, where $V_{l4}$ is the mixing between the light and heavy neutrinos. 
The decay length of heavy neutrino $d_{LR}$ in the mLRSM for $M_{W_R}\sim 10$~TeV and the same mass of the heavy neutrino is typically larger than $d_N$ by a factor of $2-3$.
This larger displaced decay length could be used in the analysis of Ref.~\cite{Cottin:2018nms} to improve the efficiency of displaced-vertex cuts and, therefore, the overall efficiency.  The detailed analysis is beyond the scope of this work. For purposes of illustration and to set an optimistic potential future goal, we choose 
$\epsilon_{\text{LLP}}^{\text{LHC}}\ \epsilon_{\text{prompt}}^{\text{LHC}} = 30\%$ as a benchmark value that one might consider as an ultimate efficiency target. 
within the mLRSM. 
With this optimistic efficiency $\epsilon_{\text{LLP}}^{\text{LHC}}\ \epsilon_{\text{prompt}}^{\text{LHC}} = 30\%$, a large portion of parameter space in the reach of future ton-scale $\onbb$-decay experiments can be probed in the LLP searches at the HL-LHC main detectors ATLAS/CMS in case of only one heavy neutrino as shown in Fig.~\ref{LLP_LHC_b}. Besides, the reach to $M_{W_R}$ mass can be extended to $\sim 13\tev$.

 Until now, we only consider the maximal value for $\tan\beta=0.3$. It is also interesting to ask how the magnitude of left-right mixing can affect the interplay of LLP searches and $\onbb$-decay searches.
To see it, in Fig.~\ref{LHC_several} we show the $\tan\beta$ dependence of (a) the reaches in the LLP searches at the HL-LHC main detectors ATLAS/CMS with the overall efficiency of $30\%$ and (b) current $\onbb$-decay searches. Notice that the process we propose as shown in Fig.~\ref{fig:LHC_signal_process} demands a non-zero $\tan\beta$. 
But this is not necessary for $\onbb$ decay. In the limit of $\tan\beta\rightarrow 0$, 
there are still non-vanishing contributions to $\onbb$-decay rate from the exchange of two $W_R$ bosons and an intermediate $N$, as well as the exchange of two $W_L$ bosons and intermediate light neutrinos, which are included in Eq.~\eqref{eq:half-life}. From Fig.~\ref{LHC_several}, we find that with the overall efficiency of $30\%$, the LLP searches at HL-LHC can probe the region of $M_{W_R}$ and $m_N$ inaccessible to current $\onbb$-decay searches when  $\tan\beta\gtrsim 0.1$.

\begin{figure}
 \centering
 \includegraphics[width=0.9\columnwidth]{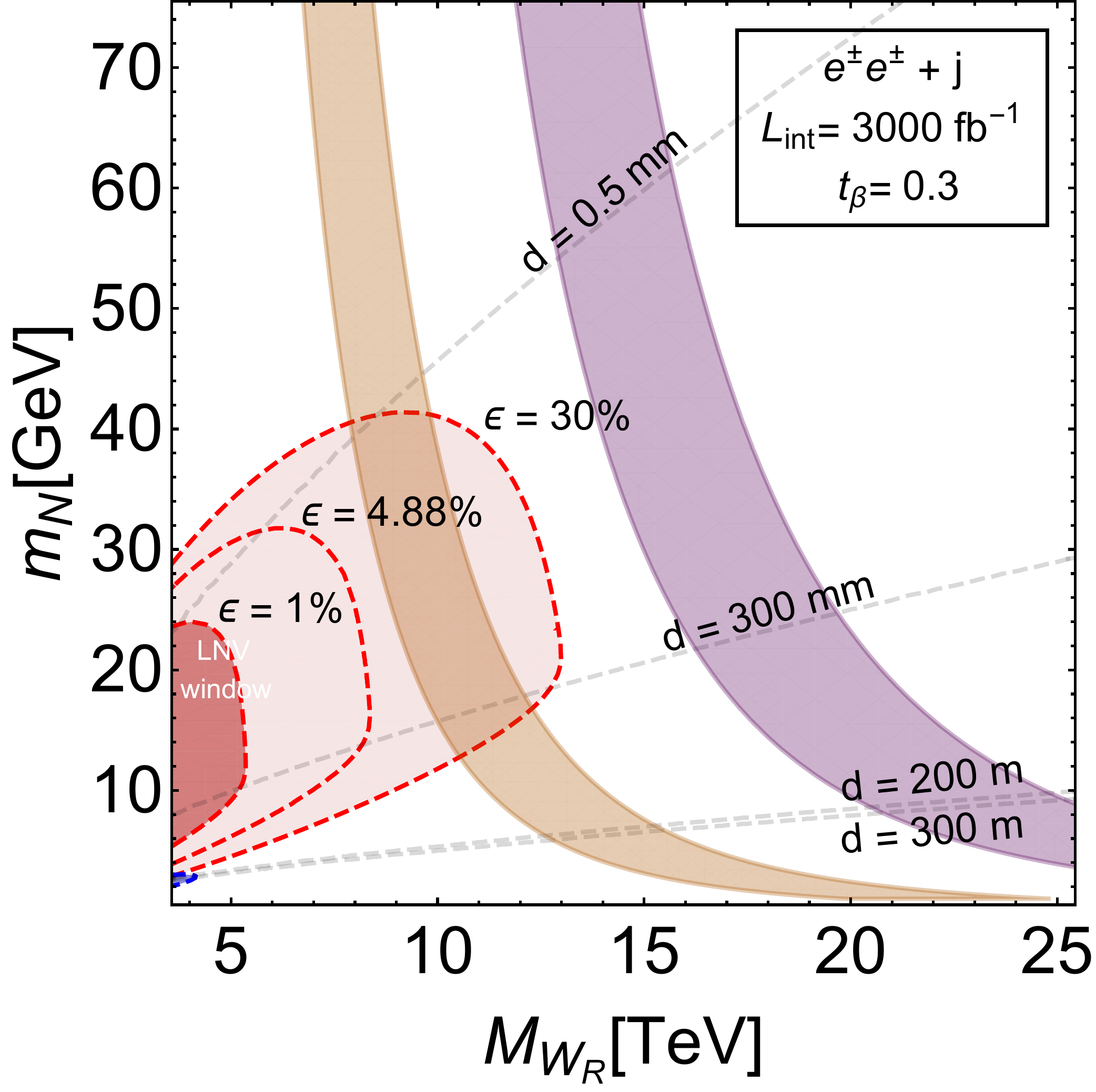} 
 \caption{ Comparison of the reaches in the LLP searches featuring LNV at the HL-LHC main detectors ATLAS/CMS and MATHUSLA and $\onbb$ decay. The MATHUSLA reach is estimated using Eq.~\eqref{NMATHUSLA}, and it is represented in the dark blue region. The dashed gray lines describe the boosted decay length of heavy neutrino. The sensitivities of  LLP searches at the HL-LHC main detectors ATLAS/CMS and $\onbb$ decay are the same as Fig.~\ref{LLP_LHC}. 
}
 \label{LLP_MAT}
\end{figure}
%


Finally, in Fig.~\ref{LLP_MAT} we  show the reaches at the HL-LHC main detectors ATLAS/CMS and MATHUSLA,
together with the $\onbb$-decay exclusion regions in the $(M_{W_R},m_N)$ plane for $\tan\beta=0.3$. The $2\sigma$ exclusion limit at MATHUSLA is obtained by requiring $N_{\text{obs}}^{\text{MATH}}=3$ assuming zero background and $\epsilon_{\text{LLP}}^{\text{MATH}}=1$. 
We can see that the search at the MATHUSLA detector could probe the region of heavy neutrino masses between $m_N = (1,4)$ GeV and $M_{W_R}$ below about 5~TeV.  Current $\onbb$-decay searches rule out this portion of the parameter space -- assuming no cancellations between different heavy neutrino contributions to the $\onbb$-decay rate. 

\section{Conclusions}\label{secV}

In the context of the mLRSM with a non-zero $W_L-W_R$ mixing, we have studied the complementarity between current and future ton-scale $\onbb$-decay searches and long-lived particle searches at the high-luminosity LHC  main detectors ATLAS/CMS and the proposed MATHUSLA detector.  In contrast to previous studies,
we have shown that the cross-section for heavy neutrino production channel $pp\rightarrow W^{\pm}\rightarrow e^{\pm}N$  may be larger than that for the production $pp\rightarrow W_R^{\pm}\rightarrow e^{\pm}N$ (KS process) for $W_R$ boson mass above 5 TeV and heavy neutrino masses $m_N\leq M_W$. Our work motivates new experimental searches for heavy Majorana neutrinos in the same-sign dilepton channel with longer decay lengths with respect to the SM augmented by three sterile heavy neutrinos. To compete with future ton-scale $\onbb$-decay mass reach,  LHC analysis needs to be improved with more delicate displaced-vertex cuts, such as the one proposed in Ref.~\cite{Cottin:2018nms}. These improved LHC searches can then be  used to probe the $W_L-W_R$ mixing comparable to and even better than current capabilities and to achieve higher mass reach relative to the standard KS process. We emphasize that the LLP searches we propose complement $\onbb$ decay searches. 

We designed two search strategies at the LHC depending on whether LNV  in the final state is manifest or not. 
In the first strategy corresponding to the LNV signal region, we require two same-sign electrons and at least one jet. In the second strategy, which does not feature LNV, we require at least one electron and one jet. We also identified the primary sources of background events for the same-sign dilepton channel.  For heavy neutrino masses below the electroweak scale, the $W_R$  mass reach at the high-luminosity LHC main detectors ATLAS/CMS may extend up to $\sim 13$ TeV for non-zero $W_L-W_R$ mixing with $\tan\beta=0.3$. Finally, we show that current $\onbb$-decay constraints already rule out the portion of the parameter space that the MATHUSLA detector would probe -- assuming no accidental cancellations in the decay rate.

\section*{Acknowledgements.}
JCV was supported in part under the US Department of Energy contract DE-SC0015376. GL, MJRM, and JCV were partially funded under the US Department of Energy contract DE-SC0011095. MJRM was also supported in part under National Science Foundation of China grant No. 19Z103010239.

\appendix
\section{Selection efficiencies}\label{appendix}
\begin{widetext}
In this appendix, we present the tables of the background selection efficiencies after the cuts mentioned in Sec.~\ref{secIII}. 

\begin{table}[h]
\centering
\tabcolsep=0.05cm
\begin{tabular}{ c | c c c c c c c c c|}
\hline
\hline
\cellcolor{nicegreen} LNV &  & \multicolumn{4}{c}{Background efficiencies}& &  \\
 \hline 
          $\sqrt{s}=$13TeV & \cellcolor{mu} $WZj$ & \cellcolor{mu} $ZZj$  &\cellcolor{mu}  $4j$ 
          &\cellcolor{mu}  $t\overline{t}j$ &\cellcolor{mu}  $W3j$ &\cellcolor{mu}  $Wj$ &\cellcolor{mu}  $Z3j$  & \cellcolor{mu} $tj$ \\
\hline
\hline
$e^\pm e^\pm j$ (b-veto)            & $3\times10^{-4}$  & $8\times10^{-5}$  & 0.0 & $4\times10^{-5}$  & $6\times10^{-5}$  & 0.0 & $2\times10^{-5}$  & $2\times10^{-5}$   \\
$\slashed{E}_T$ & $1\times10^{-4}$ & $7\times10^{-5}$ & 0.0 & $9\times10^{-6}$  & 0.0 & 0.0 & $1\times10^{-5}$ & $3\times10^{-6}$ \\
${p}_T(e_1)$   & $5\times10^{-5}$ & $4\times10^{-5}$ & 0.0 & $2\times10^{-6}$ & 0.0 & 0.0 & $8\times10^{-6}$ & $1\times10^{-6}$ \\
$M_T(e_i, \slashed{E}_T) $ & $5\times10^{-5}$ & $3\times10^{-5}$ & 0.0 & $2\times10^{-6}$ & 0.0 & 0.0 & $8\times10^{-6}$ & $1\times10^{-6}$ \\
$M(e_1,e_2)$, $M(e_i, \slashed{E}_T)$  & $3\times10^{-5}$& $3\times10^{-5}$ & 0.0 & $2\times10^{-6}$ & 0.0 & 0.0 & $7\times10^{-6}$ & $1\times10^{-6}$ \\
$l_T$, $d_{xy}$ & 0     & 0 &  0 &  0 &  0 & 0 &  0 &  0
   \\ \hline
\end{tabular}
\caption{SM background processes at 13 TeV in the LNV signal region discussed in Sec.~\ref{secIV}.
}
\label{tabBckgLNV}
\end{table}
%
\begin{table}[h]
\centering
\tabcolsep=0.05cm
\begin{tabular}{ c | c c c c c c c c c c c|}
\hline
\hline
\cellcolor{lightgreen} LNC  &  & \multicolumn{4}{c}{Background efficiencies}& &  \\
 \hline 
          $\sqrt{s}=$13TeV &\cellcolor{mu}   $WZj$ & \cellcolor{mu} $ZZj$  &\cellcolor{mu}  $4j$ 
          &\cellcolor{mu}  $t\overline{t}j$ &\cellcolor{mu}  $W3j$ &\cellcolor{mu}  $Wj$ & \cellcolor{mu} $Z3j$  & \cellcolor{mu} $tj$ \\
\hline
\hline
$e^\pm j$ (b-veto)            & $3\times10^{-2}$  & $1\times10^{-2}$  & $7\times10^{-4}$ & $6\times10^{-3}$  & $3\times10^{-2}$  & $1\times10^{-2}$  & $8\times10^{-3}$  & $5\times10^{-3}$   \\
$\slashed{E}_T$ & $1\times10^{-2}$ & $8\times10^{-3}$ & $6\times10^{-4}$ & $1\times10^{-3}$  & $1\times10^{-2}$ & 0 & $7\times10^{-3}$ & $1\times10^{-3}$ \\
${p}_T(e)$   & $8\times10^{-3}$ & $5\times10^{-3}$ & $5\times10^{-4}$ & $4\times10^{-4}$ & $5\times10^{-3}$ & 0 & $4\times10^{-3}$ & $6\times10^{-4}$ \\
$M_T(e, \slashed{E}_T) $ & $6\times10^{-3}$ & $5\times10^{-3}$ & $5\times10^{-4}$ & $3\times10^{-4}$ & $3\times10^{-3}$ & 0 & $4\times10^{-3}$ & $4\times10^{-3}$ \\
$M(e, \slashed{E}_T)$  & $5\times10^{-3}$& $4\times10^{-3}$ & $5\times10^{-4}$ & $2\times10^{-4}$ & $2\times10^{-3}$ & 0 & $3\times10^{-3}$ & $3\times10^{-4}$ \\
$d_{xy}$ & 0    & 0 &  0 &  0 &  0 & 0 &  0 &  0
   \\ \hline
\end{tabular}
\caption{SM background processes at 13 TeV in the LNC signal region discussed in Sec.~\ref{secIV}.  
}
\label{tabBckgLNC}
\end{table}
\end{widetext}

\bibliographystyle{jhep}
\bibliography{bibliography}
\end{document}